\documentclass[12pt, twoside]{article}

\usepackage{amsfonts}
\newcommand{\bb}{\begin {minipage} {3cm}\begin{center}}
\newcommand{\ee}{\end{center}\end{minipage}}
\newcommand{\bc}{\begin {minipage} {2.5cm}\begin{center}}
\newcommand{\bd}{\end{center}\end{minipage}}

\newcommand {\q}{\begin{quote} \small}

\newcommand {\be}{\begin{equation}}
\newcommand {\e}{\end{equation}}
\newcommand {\bea}{\begin{eqnarray}}
\newcommand {\ea}{\end{eqnarray}}
\newcommand {\g}{{\mathfrak g}}

\newcommand {\fract}[2]{\mbox{${\textstyle{\frac{#1}{#2}}}$}}

\begin{document}
\newtheorem {lemma}{Lemma}[subsection]
\newtheorem {theorem}{Theorem}[subsection]
\newtheorem {coro}{Corollary}[subsection]
\newtheorem {defi}{Definition}[subsection]
\newtheorem {obs}{Remark}[subsection]
\newtheorem {prop}{Proposition}[subsection]
\newtheorem {exa} {Example} [subsection]

\begin{flushright}
FTUV-00-0627 , IFIC-02-23 \\ 
DAMTP-2000-18\\
\vskip .5cM
27th June, 2000
\end{flushright}
\vskip 1cm
\begin{center}
{\Large {\bf Compilation of relations for the antisymmetric}} 
\\\vspace{0.1cm}
{\Large {\bf tensors defined by the Lie algebra cocycles of $su(n)$}}\\
\vspace{1cm}

\begin{sl}
{\large J.A. de Azc\'arraga$^1$ and A.J. Macfarlane$^2$}\\
\vskip .5cm
$^1$Dpto. de F\'{\i}sica Te\'orica and IFIC, Facultad de Ciencias,\\
46100-Burjassot (Valencia), Spain\\
$^2$Centre for Mathematical Sciences, D.A.M.T.P.,\\
Wilberforce Road, Cambridge CB3 0WA, UK \\
\vskip 1cm
\end{sl}
\end{center}
\begin{abstract}

This paper attempts to provide a comprehensive compilation of results, many 
new here, involving the invariant totally antisymmetric tensors 
(Omega tensors) which define the Lie algebra cohomology cocycles of 
$su(n)$, and that play an essential role in the
optimal definition of Racah-Casimir operators of $su(n)$. 
Since the Omega tensors occur naturally within the algebra of totally 
antisymmetrised products of $\lambda$-matrices of $su(n)$, relations within 
this algebra are studied in detail, and then employed to provide 
a powerful means of 
deriving important Omega tensor/cocycle identities.
The results include formulas for the squares of all the Omega tensors of 
$su(n)$. Various key derivations are given to illustrate the methods employed.
\end{abstract}

\pagebreak

\tableofcontents

\section{Introduction}

A recent paper \cite{tensors} gives a systematic account of the 
invariant symmetric and skewsymmetric primitive tensors 
that may be constructed on a compact simple 
Lie algebra $\g$ of rank $l$. The new family of symmetric tensors 
introduced in \cite{tensors} allows the direct construction of the 
$l$ primitive Racah-Casimir operators for $\g$; the antisymmetric
tensors determine the $l$ primitive Lie algebra cohomology cocycles of $\g$.   
We refer {\it e.g}, to \cite{tensors,dApb} and references therein 
for the definitions and explanation of the significance of the 
invariant skewsymmetric tensors
\be 
\label{A1} 
{\Omega^{(2m-1)}}_{i_1 \, i_2 \,\dots \, i_{2m-1}} \equiv
\Omega_{i_1 \, i_2 \,\dots \, i_{2m-1}} \quad (i_1,  \, i_2 \, \dots \in 
\{1,2, \dots ,{\rm dim}\,\g \})\quad ,
\e 
\noindent
that are associated with the primitive cocycles of $\g$, 
of $l$ different orders $q=2m-1$, where $m$ is the order of the 
associated Racah-Casimir operators. 
The $l$ allowed values of $q$ for the different
Lie algebra cohomology groups of each $\g$ are well-known (see {\it e.g.} 
\cite{tensors,dApb} for tables and further references). For $su(n)=A_l,
\; l=n-1$, $m=2,3,\dots,n$, the cocycles $\Omega^{(2m-1)}$ have orders
$q=3, 5,  \dots , 2m-1$ 
and indices $i_1,\dots,i_{(2m-1)}=1,\dots,n^2-1$.  
Each tensor (\ref{A1}) is used in \cite{tensors} to
define exactly one member of a family of maximally traceless (Sec.2)
fully symmetric tensors $t$. These have a very favourable status within 
the set of totally symmetric tensors, because each of them can be 
used to define one {\it primitive} Racah-Casimir operator for $\g$, 
$l$ in all, and no more. For example, if one specialises 
to $su(3)$ the general $su(n)$ definition of an $m$-th order 
tensor $t^{(m)}$ of our 
favoured family of symmetric tensors, one finds that the definition collapses 
to zero for any $m>3$, in keeping with the fact that $su(3)$ has 
no primitive Racah-Casimir operators of any order greater than three. 

These matters are fully discussed and illustrated in 
\cite{tensors,dAMcas}, 
which pay especial attention to the case of 
the $SU(n)$ group. 
However, if one wants to make application of tensors 
like (\ref{A1}), for example, in the construction of higher supercharges 
\cite{dAM2000} 
and in the quantum mechanics of particles with $SU(n)$ colour,
or in the definition of BRST-like operators and 
higher order Hodge analysis \cite{cdAmpb}, 
one finds the need for more identities involving them than are to 
be found in \cite{tensors} or elsewhere. 
There are also other areas in which good control over the properties of the 
Omega-tensors/cocycles (\ref{A1}) is valuable. One of these is
in the discussion of multi-bracket generalisations of Lie algebras 
\cite{dApb} and higher order linear Poisson structures of the 
type introduced in \cite{jaappb}. Another is in the construction of 
Wess-Zumino terms for effective actions in space-times of various 
dimensions (see \cite{dHw,dH,effacts} and references therein) and, 
in general, in the group theory factors that may appear in particle 
physics. A recent study of this last subject is \cite{vRSV}.

The aim in this paper is to
collect all that we currently know regarding the properties of and the 
identities involving the $su(n)$ algebra skewsymmetric Omega 
tensors (see (\ref{B4})--(\ref{B6}) below).
Our approach divides itself into three stages. The first presents a 
discussion of the Omega tensors of $su(n)$ that sets out from 
their definitions, and utilises only the properties of the $f$- 
and $d$-tensors \cite{msw,tensors} to deliver its output,
which is then employed in the second stage. This stage is based 
on the fact that the Omega tensors play a central role in the 
discussion of the algebra of totally antisymmetrised products of  
an even or odd number of lambda-matrices of $su(n)$ 
\be \label{A2} 
\lambda_{[ijk \dots s]}= \lambda_{[i} \lambda_j \lambda_k
\dots \lambda_{s]} \quad , \quad i,j,\dots,s= 1,\dots, n^2-1 \quad,
\e \noindent
and accordingly we provide an extensive discussion of results within this 
algebra. The completeness properties and trace identities for such products
thus  obtained then
give rise to the powerful approach contained in the third stage of our
programme, which allows us to derive, amongst other results, one of special
importance. Defining the fully contracted scalar
\be 
\label{A3} 
(\Omega^{(2s+1)}){}^2=\Omega_{i_1 j_1 \dots i_s j_s k} \,
                    \Omega_{i_1 j_1 \dots i_s j_s k} \quad , \e \noindent
we derive the recursion relation (\ref{S13})
\be \label{A4}
(\Omega^{(2s+1)}){}^2=\fract{4}{2s(2s-1)}\, (n^2-s^2) \,
(\Omega^{(2s-1)}){}^2 \quad , \e
\noindent 
and its explicit solution (\ref{S14}).

The content of this paper is organised as follows.
Sec. 2  gives the basic definitions of Omega tensors. Then Sec. 3 
gives results, within the first stage of our study, classified 
roughly according to types, {\it e.g.} Jacobi or $Ad$-invariance results,
contractions including those 
with the structure constants $f_{ijk}$ of $su(n)$
(that define the three-cocycle $\Omega^{(3)}$), recursive 
relations, duality relations, product relations, uniqueness 
questions for antisymmetric tensors, 
{\it etc}. Some derivations 
(of formulas not derived elsewhere) are
given in Sec. 4  in order to illustrate methods  employed within 
stage one of our work. In Sec. 5, we turn to the development of the 
algebra of the quantities (\ref{A2}), 
using results for Omega tensors deduced independently of
it. This then enables the attack in Sec. 6 on the identities (\ref{A3}) and
(\ref{A4}), using completeness and trace properties of the products (\ref{A2})
found in Sec. 5. Within this approach some auxiliary results are 
merely quoted in Sec. 6, and proved in Sec. 7. Some critical questions raised 
within the discussion in Sec. 3.7 of the uniqueness of Omega tensors are
answered in Sec. 8, using lambda-matrix methods.

There have been in the literature very many discussions of
the invaritensors, symmetric and skewsymmetric, associated with a simple 
Lie algebra $\g$ of rank $l$. A significant recent one is \cite{vRSV};
lists of references are given in \cite{tensors,dApb}. 
However such studies often do not consider the full set 
of invariant tensors for $\g$, neglecting the $(l-1)$ higher order 
Lie algebra cocycles {\it i.e.}, the invariant higher order
antisymmetric Omega tensors. Our paper \cite{dAMcas}
emphasises the crucial role these Omega tensors play  not only in the method 
we advocate to define Racah-Casimir operators, but also in our discussion 
of their eigenvalues and the corresponding generalised Dynkin indices. One 
additional motivation for the present paper
is to make readily available a comprehensive listing of  
results involving Omega tensors
that  are needed  for that programme. 

\section{Definitions of $su(n)$ $d, \Omega$ and $t$ tensors}

We start with a family of
symmetric invariant  tensors, the $d$-family. It is easily defined 
recursively \cite{sud} starting from the standard Gell-Mann totally 
symmetric tensor $d_{ijk}$ (see eq. (\ref{E2})). First, 
one constructs
\be 
\label{B1} d^{(r+1)}{}_{{i_1} \, \dots \, {i_{r+1}}}= 
     d^{(r)}{}_{{i_1} \, \dots \, {i_{r-1}}j} \, d^{(3)}{}_{j{i_r}
{i_{r+1}}} \quad , \quad r=3,4, \dots \quad . 
\e \noindent
For $r \geq 3$, eq. (\ref{B1}) does not define totally symmetric 
tensors. The $d$-family of symmetric tensors is obtained by 
symmetrising over all free indices in (\ref{B1}) and hence is 
defined by
\be 
\label{B2}  d^{(r)}{}_{({i_1} \, \dots \, {i_r})} \quad , 
\e \noindent
where the round brackets indicate symmetrisation with
 unit weight over the set of indices enclosed. This should be done 
as economically as possible, {\it e.g.}
\be \label{B3} 
d^{(4)}{}_{(ijkl)}= \fract{1}{3}(d_{ijt}d_{klt}+
d_{jkt} d_{ilt}+d_{kit}d_{jlt}) \quad . 
\e \noindent
The lowest order symmetric tensor, the Cartan-Killing metric (since
$\g$ is compact and the generators hermitian it will be
taken as the unity) may be viewed as the 
order two member of the $d$-family (\ref{B2}), $d_{ij} \equiv \delta_{ij}$.
Since the iteration process (\ref{B1}),(\ref{B2}) can go on idefinitely, 
it is clear that not all tensors of the $d$-family are  primitive, 
since for a simple algebra $\g$ of rank $l$ there are only $l$ invariant 
primitive symmetric tensors (or, equivalently, $l$ primitive 
Racah-Casimir operators).

We now turn to the totally antisymmetric 
Omega tensors (\ref{A1}), referring to
\cite{tensors,dApb} for an explanation of their cohomological origin.
Thus we define 

\bea 
\Omega^{(3)}{}_{ijk}\equiv
f_{ijk} & = & f^a{}_{ij}\, d_{ak} \quad , \label{Bf} \\
\Omega^{(5)}{}_{ijklm}\equiv
\Omega_{ijklm} & = & f^a{}_{[ij} f^b{}_{kl]}\, d_{abm} \quad , \label{B4} \\
\Omega^{(7)}{}_{ijklmpq}\equiv
\Omega_{ijklmpq} & = & f^a{}_{[ij} f^b{}_{kl} f^c{}_{mp]} d^{(4)}{}_{(abcq)} 
\quad , \label{B5} \\
\Omega^{(9)}{}_{ijklmpqrs}\equiv
\Omega_{ijklmpqrs} & = & f^a{}_{[ij} f^b{}_{kl} f^c{}_{mp} f^d{}_{qr]} 
d^{(5)}{}_{(abcds)} \quad , \label{B5A} \\
\Omega^{(11)}{}_{ijklmpqrsuv}\equiv
\Omega_{ijklmpqrsuv} & = & f^a{}_{[ij} f^b{}_{kl} f^c{}_{mp} f^d{}_{qr}  f^e{}_{su]}
d^{(6)}{}_{(abcdev)} 
\quad , \label{B6}  
\ea 
\noindent
and so on. Here, the square brackets imply total unit weight 
antisymmetrisation over the set of indices which they enclose.
The raising of indices is trivial from a metric point of view, and is usually
used in this paper in order to exempt certain indices from the 
antisymmetrisation (or symmetrisation) effect of the square
(round) brackets.
We note that the $\Omega^{(2m-1)}$ tensor is fully skewsymmetric in its
$(2m-1)$ indices in spite of the fact that only $(2m-2)$ indices are explicitly
antisymmetrised in the r.h.s. above \cite{jaappb,tensors}.
Some further discussion of the properties of the $d$-tensors and their role 
in the definitions of the Omega tensors is given below as Sec. 2.1.

Next we use the Omega tensors to define a second family of invariant
fully symmetric tensors, the $t$-tensors, as follows \cite{tensors}
\bea 
\Omega_{ijm} f_{ija} & = &  t^{(2)}{}_{am}  \quad , \label{B6A} \\
\Omega_{ijklm} f_{ija} f_{klb} &= & t^{(3)}{}_{abm}  \quad , \label{B7} \\
\Omega_{ijklmpq} f_{ija} f_{klb} f_{mpc} & = & t^{(4)}{}_{abcq}  \quad , 
\label{B8} \\
\Omega_{ijklmpqrs} f_{ija} f_{klb} f_{mpc} f_{qrd} & = & 
t^{(5)}{}_{abcds}  \quad , \label{B9}
 \ea 
\noindent
etc.; they are fully symmetric on account of the skewsymmetry of the $\Omega$'s.
The tensors on the right hand side of (\ref{B7})--(\ref{B9}) have been 
evaluated before \cite{tensors}
\footnote{Eqs. (\ref{B10}) and (\ref{B11}) above correct the overall 
factors of (6.13) and (6.14) in \cite{tensors}; the $i$ difference
in \cite{tensors} is due to the fact that here we take the 
generators of $g$ hermitian, see (\ref{E14}).}.
We give below the expression of the lower order
$su(n)$ $t$ tensors in terms of members of the $d$-family,
\bea 
t^{(2)}{}_{ij} & = & n \delta_{ij} \quad , \label{B9A} \\
t^{(3)}{}_{ijk} & = & \fract{1}{3} n^2 d_{ijk} \quad , \label{B10} \\
 t^{(4)}{}_{ijkl} & = & \fract{1}{15}(n(n^2+1) d^{(4)}{}_{(ijkl)}
-2(n^2-4) \delta_{(ij}\delta_{kl)}) \quad , \label{B11} \\
 t^{(5)}{}_{ijklm} & = & \lambda(n) \left( n(n^2+5) d^{(5)}{}_{(ijklm)}
-2(3n^2-20) d_{(ijk}\delta_{lm)} \right)   \quad , \label{B12} \ea \noindent
where the function $\lambda(n)$, not determined in \cite{tensors},
turns out from the work of Sec. 6 to be given by
\be \label{B12A}
\lambda(n)= \fract{n}{105} \quad . 
\e \noindent
The tensors (\ref{B10}) and higher collapse to zero when their order 
$m$ is larger than $n$. While Eq. (\ref{B12}) can be indeed be used 
as it stands (and will be to avoid circularity of argumentation 
below), much of the information we need will be seen to follow 
from the definitions (\ref{B7})--(\ref{B9}) and the properties of Omega 
tensors.

The $t$-tensors are totally symmetric and, unlike the higher
($m>3$) order $d$-tensors, they are 
orthogonal to all other $t$-tensors of different order
(Lemma 3.3 of \cite{tensors}). For instance, for $t^{(4)}$ 
this means
\be \label{B13} 
t^{(4)}{}_{ijkl} \delta_{ij}=0 \quad , \quad
t^{(4)}{}_{ijkl} t^{(3)}{}_{ijk}=0  
\quad . 
\e \noindent
In contrast, since trace formulas for $d$-tensors easily give
\be \label{B14} 
d^{(4)}{}_{(ijkl)} d_{ijm}=\fract{2}{3}  
\fract{(n^2-8)}{n} d_{klm} \quad , 
\e \noindent
the contraction of only two indices gives
\be \label{B14A} 
t^{(4)}{}_{ijkl} t^{(3)}{}_{ijm}=\fract{1}{3} \,n^2 t^{(4)}{}_{ijkl} d_{ijm}=
\fract{2}{45} n^2 \, (n^2-9) t^{(3)}{}_{klm} \quad .
\e 
\noindent
For $t^{(5)}$ we have 
\be \label{B15} 
t^{(5)}{}_{ijklm} \delta_{ij}=0 \quad , \quad
t^{(5)}{}_{ijklm} t^{(3)}{}_{ijk}=0  \quad , \quad
t^{(5)}{}_{ijklm} t^{(4)}{}_{ijkl}=0   
\e \noindent
or, in other words, the maximal contraction of the indices   of
two $t$-tensors of {\it different} order is zero.

Another issue concerns the claims made above that $t^{(4)}$ vanishes 
identically for
$n=3$, and that $t^{(5)}$  vanishes identically for 
$n=3$ and $n=4$. This is a point one will see clearly 
illustrated in many places,
and is the result of factors becoming zero or of relations expressing
the $d^{(m)}$ tensors in terms of primitive ones when $m>n$.
The necessary identities special to $su(3)$ and 
$su(4)$ are noted in \cite{tensors}. Indeed almost all we need here in the way
of identities involving the $d$- and $f$-tensors of $su(n)$ is presented 
in \cite{tensors}, especially in the appendix. See also 
\cite{msw} and \cite{MPf}. 

\subsection{More about $d$-tensors}

Below we shall need the identities that express the $Ad$-invariance 
of the $d$-tensors,
namely,
\bea
d_{t(ij} f_{k)st}  & =& 0 \quad , \label{Y1} \\
{d^{(4)}}_{t(ijk} f_{l)st}  & =& 0 \quad , \label{Y2} \\
{d^{(5)}}_{t(ijkl} f_{m)st}  & =& 0 \quad , \label{Y3} \ea \noindent
and so on.

One use of (\ref{Y1}) is as follows. 
Referring first to (\ref{B4}), we note that the symmetry 
properties of the $SU(n)$-invariant $d$- and $f$-tensors permit 
the left-hand square bracket to be moved, without 
altering the definition, one place to the right, so as 
to enclose only $j,k$ and $l$. Then application of (\ref{Y1}) allows one to
show that the right side of (\ref{B4}) is antisymmetric 
under the interchange of $i$ and $m$, and hence, as mentioned, indeed 
defines a totally antisymmetric quantity.

Referring next to (\ref{B5}), we note that it may be 
simplified in either of two ways but not simultaneously both: one of these 
is the move of the left hand  square bracket one place to the right 
(or alternatively the right hand one to the left), the other 
employs the result
\be 
\label{Y4}
d^{(4)}{}_{(abcq)}=d^{(4)}{}_{(abc)q} \quad , \e \noindent
and thereby allows 
$d^{(4)}{}_{(abcq)}$ to be replaced in (\ref{B5}) by one of the terms of 
$d^{(4)}{}_{(abc)q}$, {\it e.g.}
$d_{xab} d_{cqx}$. It is the use of (\ref{Y4}) that deserves 
close attention. It displays a simplifying feature of the $d^{(4)}$ 
situation that does
not generalise systematically to $d^{(r)}$ for $r>4$. For $r=5$, we have
\be \label{Y4A} 
d^{(5)}{}_{(abcdq)} =  \fract{1}{5} d^{(5)}{}_{(ab}{}^q{}_{cd)}
+ \fract{4}{5} d^{(5)}{}_{(abcd)q} \quad , \e \noindent
where
\bea 
 d^{(5)}{}_{(ab}{}^q{}_{cd)} & = & d_{(ab}{}^x d^{xqy} d^y{}_{cd)} \quad,
\label{Y6} \\
d^{(5)}{}_{(abcd)q} & = & d_{(ab}{}^x d^x{}_c{}^y d^y{}_{d)q} \label{Y7}
\quad . 
\ea \noindent 
Inspection of the evident tree-diagram representation
of the tensors occurring here makes clear the fractions seen in (\ref{Y4A}).

In (\ref{Y4A}) we meet an obstacle to extending, to (\ref{B6}) and beyond,
the simple proof that allowed 
the right hand round bracket in (\ref{B5}) to be moved one place to the left.
It is nevertheless a generally allowed step, providing a valuable 
simplification of the definitions of $\Omega^{2s+1)}$ for 
$s \geq 4$. However, to obtain a convenient proof of this, we need 
to have recourse to lambda matrix methods, and so, will return to the
matter in Sec.5.2.

Similarly
\be \label{Y8}
d^{(6)}{}_{(abcdeq)} = \fract{1}{3} d^{(6)}{}_{(ab}{}^q{}_{cde)}
+ \fract{2}{3} d^{(6)}{}_{(abcde)q} \quad . 
\e 
\noindent 
The fractions in the RHS of (\ref{Y8}) arise because  
the tree diagrams in use are 
trees with four equivalent end twigs and two equivalent non-end twigs.

An additional complication enters for symmetric tensors 
of order six, one that has already been observed 
in \cite{tensors}. The tensor $d^{(6)}$ that enters 
the definition (\ref{B5A}) of $\Omega^{(11)}$ is the $r=6$ member
of the family (\ref{B2}). But it is not the only primitive symmetric sixth 
order tensor that can be defined. One also has $d^{(6)\prime}$ given by
\be \label{Y9}
d^{(6)\prime}{}_{(abcdef)}=d_{(ab}{}^x d_{cd}{}^y d_{ef)}{}^z d_{xyz} \quad .
\e \noindent 
The tensors $d^{(6)}{}_{(abcdef)}$ and $d^{(6)\prime}{}_{(abcdef)}$
are for our purposes equivalent. 
It is shown in \cite{tensors} (below eq. (A.21) there) 
that they differ by non-primitive terms
which are symmetrised products of lower order $d$-tensors. 
The claimed equivalence follows from the fact that such 
non-primitive terms cannot contribute to
the definition (\ref{B5A}) of $\Omega^{(11)}$ because of Jacobi identities.

Inspection of the relevant tree diagram shows that
\be \label{Y10} 
d^{(6)\prime}{}_{(abcdeq)}=d^{(6)\prime}{}_{(abcde)q}
\quad . 
\e \noindent

\section{Identities involving the Omega tensors}

These are mostly displayed for Omega tensors of lower order for 
obvious reasons.
But one can often see patterns that would guide an attack on higher order 
analogues that may now seem out of reasonable reach, or perhaps just until 
the need of a specific application
provides the necessary motivation. 
The trace methods for products of 
the hermitian $D$- and $F$-matrices \cite{kr}, where 
$(D_i)_{jk}=d_{ijk}$ and $(F_i)_{jk}=-i f_{ijk}$,
such as are seen in use in
the derivations presented in Sec. 4, become discouraging 
when one cannot avoid doing a trace that is more than of
fourth order, unless one can harness computational skills like those of 
\cite{vRSV}.
Also, finding a viable path through increasing complication becomes 
progressively more taxing. The necessity for going on 
in later sections to develop an alternative approach -- that which makes 
systematic use of lambda matrices --
come into evidence in this way.

\subsection{Jacobi identities}

First, we consider the Jacobi identities which 
express the $Ad$-invariance of the Omega tensors.
The $Ad$-invariance of $\Omega^{(3)}{}_{ijk}=f_{ijk}$ is expressed
by the  Jacobi identity,
\be 
\label{C2} f_{t[ij} f_{k]lt}= f_{t[ij} f_{k]ls} \delta_{st}= 0 
\quad . 
\e
\noindent 
For higher $\Omega$'s $Ad$-invariance gives
\bea \Omega_{t[ijkl} f_{p]qt} & = & 0 \label{C3} \quad, \\
    \Omega_{t[ijklpq} f_{r]st} & = & 0 \label{C4}\quad,  \\
   \Omega_{t[ijklpqrs} f_{u]vt} & = & 0 \quad .\label{C5} \ea \noindent  
In analogy with the second way of writing the Jacobi identity (\ref{C2}),
we may usefully expand (\ref{C3})--(\ref{C5}) in terms of the higher members
of the $d$-family (\ref{B2}) getting
\bea 
f^a{}_{[ij} f^b{}_{kl} f^c{}_{p]q} d_{abc}=0   \label{C6} \quad, \\
f^a{}_{[ij} f^b{}_{kl} f^c{}_{pq} f^d{}_{r]s} d^{(4)}{}_{(abcd)}=0 \quad .
\label{C7} 
\ea \noindent
In the last two results, obviously the right hand square bracket can be 
moved one place to the right. In the case of (\ref{C7}) this allows the round 
symmetrising brackets to be taken off $d^{(4)}$
since  $f^a{}_{[ij} f^b{}_{kl} f^c{}_{pq} f^d{}_{rs]}$ is fully 
symmetric in $abcd$. 

Also, using (\ref{C4}), eq. (\ref{C7}) can be rearranged to
read 
\be \label{C8} \Omega_{t[ijkl} \Omega_{pqr]st}=0 \quad ,
\e
\noindent
an expression that may be understood as the generalised Jacobi identity 
(GJI) for a higher (fourth) order multibracket algebra. In the general
case, the GJI reads \cite{dApb,jaappb}
\be 
\label{GJI}
\Omega_{t[i_1 \dots i_{2m-2}} \Omega_{i_{2m-1} \dots i_{4m-5}]i_{4m-4}t}=0  
\quad. 
\e
This is an {\it identity} that follows directly if one takes the 
coordinates of $\Omega^{(2m-1)}$ as the generalised 
structure constants (with $(2m-1)$ antisymmetric indices) of 
a multi-bracket Lie algebra of order $(2m-2)$. As in the standard case
(for $m=2$, eq. (\ref{GJI}) reduces to the JI, eq. (\ref{C2})), 
only the associativity of the $(2m-2)$ entries in the fully
skewsymmetric multibracket is required to obtain eq. (\ref{GJI}). 

The importance of the Jacobi identities can hardly be overemphasised: they
are essential to the simplification of other identities in all classes,
as will be seen below.
Certain other results which bear a close resemblance to  
(\ref{C3})--(\ref{C5}) are also valid, namely
\bea
 \Omega_{ti[jkl} f_{pq]t} & = & 0 \label{C3X} \quad,  \\
 \Omega_{ti[jklpq} f_{rs]t} & = & 0 \label{C4X} \quad,  \\
 \Omega_{ti[jklpqrs} f_{uv]t} & = & 0 \quad ,\label{C5X} 
\ea \noindent  
{\it etc}. One proves these results by inserting the definitions of the Omega
tensors into their left sides and making simple rearrangements that allow 
(\ref{C2}) to be used to produce the answers zero.
Relations of this type may be understood as {\it mixed}
GJI identities. Their general form is 
\be 
\label{MGJI}
\Omega_{t[i_1\dots i_r}\Omega_{i_{r+1} \dots i_{r+s-1}] i_{r+s} t}=0
\quad (r \;\; {\rm and}\;\; s \;\; {\rm even}) \quad,
\e
and constitute consistency relations that must be satisfied by
the generalised structure constants, and hence have the same origin 
as the GJI (see \cite{aipb}). Here we have followed the converse
path, showing that these relations follow from the 
definition of the $\Omega$ tensors.

It is also worth noting that 
many of the results of this section can be identified as cases of 
the general result Lemma 3.1 of \cite{dApb}:
\be \label{L3.1}
f^{p_1}{}_{[i_1 j_1} \cdots f^{p_s}{}_{i_s j_s]} k_{(p_1 \cdots p_s)} =0
\quad , \e \noindent
where $k_{(p_1 \cdots p_s)}$ is any $Ad$-invariant totally 
symmetric tensor of order $s$.

\subsection{On the definition of the $\Omega$ tensors}

Since we have introduced the Omega tensors using the recursively 
defined $d$-tensors (\ref{B2}), and then used the Omega tensors to obtain 
the preferred family of $t$-tensors (eqs.(\ref{B6A})--(\ref{B9}), 
(\ref{B9A})--(\ref{B12}) etc.), one might
well ask why we did not need the latter in order to start the process off in 
the first place. The answer is that the
non-primitive  product terms that appear as 
the tails of the $t$-tensors cannot
contribute to the Omega tensors at all in virtue of Jacobi identities of the 
type given in 
Sec. 3.1. In fact, non-primitive invariant symmetric tensors 
do not contribute to the $\Omega$ tensors, making the definitions 
(\ref{B4})--(\ref{B6}) unique
(see \cite{tensors}, Cor. 3.1).
For example ({\it cf.} (\ref{B5})), 
there is no need to 
contemplate a contribution to $\Omega^{(7)}$ proportional to
\be 
\label{C9} 
f^a{}_{[ij} f^b{}_{kl} f^c{}_{mp]} \delta_{(ab} \delta_{cq)}
\quad , \e 
\noindent 
since it vanishes by eq. (\ref{C2}). To see that a similar state of 
affairs applies to a putative contribution to $\Omega^{(9)}$ like
\be \label{C9A} f^a{}_{[ij} f^b{}_{kl} f^c{}_{mp} f^d{}_{qr]} d_{(abc} 
\delta_{ds)}
\quad , \e 
\noindent
requires the use of both (\ref{C2}) and (\ref{C6}), depending upon where $s$ 
occurs in the five terms of the expansion of $d_{(abc} \delta_{ds)}$.
Considerations like those described often employ steps 
like $[ijklmp \dots ]=[i[jkl]mp \dots ]$ ; unit weighted brackets 
are convenient for such use. Thus apart from overall normalisation,
replacing $d$-tensors (\ref{B2}) by $t$-tensors (see eq. (\ref{B10}))
in the definitions of the Omega tensors has of no effect
since, by virtue of (\ref{L3.1}), the non-primitive parts
in which they differ do not contribute.

\subsection{Recursive identities}

We note  the important results relating $\Omega^{(5)}$ to
$\Omega^{(7)}$ and  $\Omega^{(7)}$ to $\Omega^{(9)}$ respectively
\bea  \Omega^{(7)}{}_{ijklmpq} & = & \Omega^{(5)}{}_{t[ijkl} 
f_{mp}^s d_{q]st} \quad, 
\label{C10} \\
   \Omega^{(9)}{}_{ijklmpqrs} & = & \Omega^{(7)}{}_{t[ijklmp} 
f_{qr}{}^u d_{s]ut} \label{C11}
\quad , 
\ea \noindent
which are the two lowest versions of a general result 
(\cite{tensors}, eq. (7.6)). Having written these results, 
one sees that the definition of $\Omega^{(5)}$ provides
the first member of the series, the identification
\be 
\label{C12} \Omega^{(3)}{}_{ijk}= f_{ijk} \quad,
\e \noindent 
having been noted already in (\ref{Bf}).

Each of the two results just displayed can usefully 
be presented in a different form
\bea  \Omega^{(7)}{}_{ijklmpq} & 
= & f_{x[ij} \Omega^{(5)}{}_{klmp}{}^y d_{q]xy} \quad, 
\label{C10A} \\
   \Omega^{(9)}{}_{ijklmpqrs} & = & \Omega^{(5)}{}_{x[ijkl} 
\Omega^{(5)}{}_{mpqr}{}^y d_{s]xy} \label{C11A}
\quad ;
\ea \noindent
evident generalisations may be expected to hold. It is 
easy also to use Jacobi identities to show if one replaces $d$-tensors 
by $f$-tensors on the right sides of (\ref{C10A}) and (\ref{C11A}), 
one gets the answer zero.

\subsection{Contraction of higher order Omega tensors with lower order ones}

In view of the antisymmetry of the Omega tensors, it is clear that these are
amongst the most important contractions to be considered. We find
\bea
f_{ijk} f_{ijl} & = & n \delta_{kl} = {t^{(2)}}_{kl} \quad, \label{C13} \\
\Omega_{tijkl} f_{ijs} & = & \fract{1}{2} n f_{u[kl} d_{t]us} \quad, 
\label{C14} \\
\Omega_{tijkl} f_{iju} f_{klv} & = & \fract{1}{3} n^2 d_{tuv}= t^{(3)}{}_{tuv}
\quad, \label{C15} \\
      \Omega_{tijklpq} f_{iju} f_{klv} f_{pqw} & = & 
\fract{1}{15} \left( n(n^2+1) d^{(4)}{}_{(uvwt)}-2(n^2-4)\delta_{(uv} 
\delta_{wt)} \right)= t^{(4)}{}_{tuvw}  
\label{C16} \quad . \ea \noindent
The last two of course are just the definitions of $t^{(3)}$ and $t^{(4)}$
seen from a different viewpoint. We may  
contract (\ref{C14}) further, obtaining 
\be 
\label{C17} \Omega_{tijkl} f_{ijk}=0 \quad .
\e \noindent 
It is sometimes relevant to observe the absence of the display of 
formulas that might naively be guessed
as, {\it e.g.},  for $\Omega_{tijklpq} f_{ijs}$, {\it cf.} (\ref{C14}). 
This will not reduce simply to
a multiple of $\Omega_{u[klpq} d_{t]us}$, since in this case there are other
quantities with the required symmetries available to complicate matters.
Although a useful reduction can be achieved, the result is not
clean enough to be displayed.

Families of more complicated but still useful contractions include 
the following
\bea 
\Omega^{(5)}{}_{ijkls} \Omega^{(7)}{}_{ijklpqr} & = & \fract{n}{15}(n^2-9) 
f_{u[pq} d_{r]us}
\label{K1} \\
\Omega^{(7)}{}_{ijklpqr} \Omega^{(9)}{}_{ijklpqstu} & = & \fract{2n}{105} 
(n^2-9)(n^2-16) f_{v[st} d_{u]vr} \label{K1A} \\
\Omega^{(5)}{}_{ijkls} \Omega^{(7)}{}_{ijklpqr} f_{pqt} & = & \fract{2}{45} 
n^2 \, (n^2-9)d_{rst} \label{K2} \\
\Omega^{(7)}{}_{ijklpqr} \Omega^{(9)}{}_{ijklpqstu} f_{stv} & = & 
\fract{4}{315} n^2 \, (n^2-9) (n^2-16) d_{ruv} \label{K2A} 
\ea \noindent
where factors $(n^2-9)$ and $(n^2-16)$
reflect the respective facts that $\Omega^{(7)}$ is absent
for $n=3$, and  $\Omega^{(9)}$ is absent for  $n=4$.
Eq. (\ref{K1}) can be rewritten also as
\be \label{K3} \Omega^{(5)}{}_{ijkls} \Omega^{(7)}{}_{ijklpqr} =  
\fract{2}{15}(n^2-9) f_{ijs} \Omega^{(5)}{}_{ijpqr} 
\quad , \e 
\noindent which is a recursion relation of sorts, one that be generalised 
along obvious lines, as indeed (\ref{K1}) itself has been in the 
production of (\ref{K1A}).
Eq. (\ref{K2}) is an easy consequence of (\ref{K1}), which also implies
\be 
\label{K4} \Omega^{(5)}{}_{ijklp} \Omega^{(7)}{}_{ijklpqr}=0 
\quad . \e 
\noindent 
The same applies to (\ref{K2A}) and (\ref{K1A}), so that also
\be \label{K4A} \Omega^{(7)}{}_{ijklpqs} \Omega^{(9)}{}_{ijklpqstu}=0
 \quad . \e 

Results such as  (\ref{C17}),  (\ref{K4}) and (\ref{K4A})  
are evident enough, since there is no
$SU(n)$-invariant totally antisymmetric tensor of order two. 
They point strongly towards an analogue to
the orthogonality  result for the $t$-tensors,
eqs. (\ref{B7})--(\ref{B9}), and suggest that the maximal
contraction of two Omega tensors of different order is zero.
We do not have a general proof, but as questions regarding it arise 
it will be shown that this is indeed the case.
Thus one might ask about the claim 
\be \label{K5} \Omega^{(7)}{}_{ijklpqr} f_{pqr} =0 \quad . \e 
\noindent Using the methods of this section it is indeed possible, but
not easy, to verify this by direct calculation. Alternatively, 
we may have recourse to the assertion that there exist no
$SU(n)$-invariant totally antisymmetric tensors of order four. Similarly, 
eq. (\ref{U5}) below indicates the absence of any 
$SU(n)$-invariant totally antisymmetric tensors of rank six. 
It follows that we may write
\be \label{K6} \Omega^{(9)}{}_{ijklpqrst} f_{rst} =0 \quad , \e \noindent
and similarly
\be \label{K6A} \Omega^{(11)}{}_{ijklpqrstuv} \Omega^{(5)}{}_{ijklp} =0 
\quad .
\e \noindent
Such arguments fail for the proof of
\be \label{K7} \Omega^{(11)}{}_{ijklpqrstuv} f_{tuv} =0 \quad , \e
\noindent 
because for $su(n)$ with $n>5$ (so that $\Omega^{(11)}$ exists) there
is a (non-primitive) totally antisymmetric $SU(n)$-invariant
tensor of order eight. Such matters are discussed systematically 
in Sec 3.7, the eigth-order tensor being there displayed as (\ref{U2}). 
Eq. (\ref{K7}) is nevertheless true, although we need the methods 
of later sections, lambda-matrix methods, to obtain a convenient 
approach (see Sec. 8) to its proof. Thus, as is known to be true 
for the $t$-tensors, so also it seems for the Omega tensors that 
the only invariants that can be built out of them are their fully 
contracted squares. We begin the task of evaluating these in 
the next paragraph. 

\subsection{Product identities}

We begin with the results
\bea 
f_{ijs} f_{ijt} & = & \phi_3(n) \delta_{st} \quad, \label{C18} \\
\Omega_{ijkls} \Omega_{ijklt} & = & \phi_5(n) \delta_{st} \quad, \label{C19} \\
\Omega_{ijklpqs} \Omega_{ijklpqt} & = & \phi_7(n) \delta_{st} \quad , 
\label{C20} \ea 
\noindent
that define a family of quantities of which the first few members are
\bea
\phi_3(n) & = & n \quad, \label{C21} \\
\phi_5(n) & = & \fract{1}{3} n\, (n^2-4) \quad, \label{C22} \\
\phi_7(n) & = & \fract{2}{15} \, (n^2-9) \phi_5(n) \label{C23} 
\quad . \ea \noindent
These imply the consequences
\bea
f_{ijs} f_{ijs} & = & \psi_3(n) \quad, \label{C23A} \\
\Omega_{ijkls} \Omega_{ijkls} & = & \psi_5(n) \quad, \label{C24} \\
\Omega_{ijklpqs} \Omega_{ijklpqs} & = & \psi_7(n)  \quad,  \label{C25} \ea 
\noindent 
where 
\bea
\psi_3(n) & = &  n \, (n^2-1)   \quad, \label{C25A} \\
\psi_5(n) & = &  \fract{1}{3} n \, (n^2-1)(n^2-4) \quad, \label{C26} \\
\psi_7(n) & = & \fract{2}{45} n \, (n^2-1)(n^2-4)(n^2-9)  \label{C27} 
\quad . 
\ea 
\noindent
One can speculate with confidence on the extension of these results to higher
Omega tensors. As is discussed in related contexts in \cite{tensors}, results
collapse to zero equal zero for low values of $n$, for which the relevant 
primitive cocycles do not exist. For example $su(2)$ has no cocycle of
order higher than three and $su(3)$ none higher than five. The right 
hand sides of (\ref{C26}) and (\ref{C27}) have explicit factors zero 
for the corresponding $n$-values.
The proofs of (\ref{C26}) and (\ref{C27}) are given in Sec. 4.

One often needs results more general than those so far mentioned.
We have
\bea 
\Omega_{ijkpq} \Omega_{ijkrs} & = & \fract{1}{6} \left( (n^2-6) f_{pqt} f_{rst}
+n(\delta_{pr}\delta_{qs}-\delta_{ps}\delta_{qr}) \right)  \quad, \label{C28} \\
\Omega_{ijklpqr} \Omega_{ijklpst} & = & \fract{2}{135} (n^2-9) \left(
(n^2-8)f_{qru} f_{stu}+2n(\delta_{qs}\delta_{rt}-\delta_{qt}\delta_{rs}) 
\right) \quad , \label{C29} 
\ea \noindent
which imply (\ref{C19}) and (\ref{C20}), as they should.
The right hand sides of (\ref{C28}) and (\ref{C29}) involve linear combination
of the only two fourth order tensors with the correct symmetries (so that these 
are a basis in the vector space in question). The $\Omega^{(7)}$ result
collapses for $n=3$ ($su(3)$ has no seven-cocycle) because of the explicit
factor $n^2-9$. 
Although there is no factor $(n^2-4)$ in (\ref{C28}) causing it to collapse 
for $su(2)$, the right side of (\ref{C28}) nevertheless vanishes
because, for $su(2)$, we have  $f_{ijk}=\epsilon_{ijk}$, and
$\epsilon_{ijk} \epsilon_{ipq}=\delta_{jp}\delta_{kq}-\delta_{jq}\delta_{kp}$.

\subsection{Duality results}

We give results here mainly for $su(3)$ although in principle the analysis
could be extended to higher $su(n)$. 

We mention first results involving the totally antisymmetric eighth order
epsilon tensor:
\bea
\epsilon_{ijklmnpq} \epsilon_{ijklmnpq} & = & 8! \quad, \nonumber \\
\epsilon_{ijklmnpq} \epsilon_{ijklmnpt} & = & 7! \delta_{qt} \quad, \nonumber \\
\epsilon_{ijklmnpq} \epsilon_{ijklmnst} & = & 6! (\delta_{ps} \delta_{qt}-
\delta_{pt} \delta_{qs}) \quad, \nonumber \\
\epsilon_{ijklmnpq} \epsilon_{ijklmrst} & = & 5! 3!
(\delta^{[r}_n \delta^s_p \delta^{t]}_q) \quad . \label{C30} \ea 
\noindent We note here that the factor $3!$ is present because the square 
brackets imply antisymmetrisation with  unit weight.
Next, from \cite{tensors}, we note
\bea 
12\sqrt{3} \Omega_{ijklm} & = & \epsilon_{ijklmpqr} f_{pqr} \quad ,
\label{C31} \\
20 \sqrt{3} f_{stu} & = & \epsilon_{ijklmstu} \Omega_{ijlkm}
\quad . \label{C32} \ea 
\noindent
Again results more general than the above are often called for,
as in \cite{cdAmpb}. From \cite{cdAmpb} we quote
\bea 
\epsilon_{ijklrstu} \Omega_{ijlkm} & = & 16\sqrt{3} 
\delta_{m[r} f_{stu]}  \quad ,\label{C33} \\
\epsilon_{ijklmpqr} f_{qrs} & = & 24\sqrt{3} \delta_{s[p} \Omega_{ijklm]}
\quad . \label{C34}  \ea \noindent
One may check that (\ref{C34}) implies 
(\ref{C31}), and that (\ref{C33}) implies (\ref{C32}). However to prove
(\ref{C33}) one must insert (\ref{C31}) and use an 
identity from the family (\ref{C30}). Similarly insertion of (\ref{C32})
allows proof of (\ref{C34}). Further, one may use (\ref{C30}) to show that
(\ref{C32}) follows from (\ref{C31}).
 
For an $su(4)$ result, see Sec. 8 of \cite{tensors}. Recent work of the
authors \cite{dAM2000}
actually uses duality to obtain information about $\Omega^{(9)}$ for
$su(5)$, having used MAPLE programs for data about the lower Omega tensors.

\subsection{Non-primitive antisymmetric tensors}

In Sec. 2, we defined for $su(n)$ its Omega tensors, which are a set of 
$l=(n-1)$ primitive antisymmetric tensors of orders
\be \label{U1}
3,5,7, \, \dots \, , (2n-1) \quad . \e \noindent
We have described the fundamental role they play in the discussion of primitive
Racah-Casimir operators of $su(n)$ 
(see also \cite{dAM2000})  but they are not the only antisymmetric tensors
that can be defined. One can form non-primitive,
tilded tensors ${\tilde \Omega}$, as totally 
antisymmetrised products of primitive tensors $\Omega$, {\it e.g.}

\be 
\label{U2}
{\tilde \Omega}^{(8)}{}_{ijkpqrst}=
\Omega^{(3)}{}_{[ijk}\, \Omega^{(5)}{}_{pqrst]} \quad , 
\e \noindent
which for $su(3)$ is a multiple of the eigth order $\epsilon$-tensor.
In terms of forms, eq. (\ref{U2}) determines a non-primitive de Rham cocycle
on the $SU(n)$ group manifold ({\it e.g.}, the volume form on $SU(3)$,
ignoring factors). A more interesting example arises for $su(8)$, 
$l=7$, which has seven Omega tensors of orders $3,5,7,9,11,13,15$. 
In this case, one can form a second,
tilded antisymmetric tensor of order $15$ 
\be \label{U3}
{\tilde \Omega}^{(15)}{}_{i_1 \dots i_{15}}=\Omega^{(3)}{}_{[i_1 i_2 i_3}\, 
\Omega^{(5)}{}_{i_4 \dots i_8}\, \Omega^{(7)}{}_{i_9 \dots i_{15}]} \quad , \e
\noindent 
which is non-primitive and not maximal on the $(n^2-1)=63$-dimensional 
space (manifold in the case of forms on $SU(n)$). The 
discussion of Sec. 5.1 suggests that it should have zero full
contraction with $\Omega^{(15)}{}_{i_1 \dots i_{15}}$, in virtue of results 
like (\ref{K5}).
This example illustrates a significant restriction on the
construction of non-trivial non-primitive antisymmetric tensors: 
all the primitive Omega tensors used
must have different orders.
To see this, 
notice that we may  write, {\it e.g.},
\be \label{U4}
\Omega^{(3)}= \Omega^{(3)}{}_{ijk} \, \omega^i \wedge \omega^j \wedge 
\omega^k \quad , 
\e \noindent 
where the $\omega^i$ are the left-invariant (LI) Maurer-Cartan 
one-forms on the $SU(n)$ groups manoifold, dual to the LI $su(n)$ 
generators, so that $\Omega^{(3)}$ is 
the invariant de Rham three-cocycle of coordinates $\Omega^{(3)}{}_{ijk}$. 
Obviously, $\Omega^{(3)} \wedge \Omega^{(3)}=0$ and hence
\be \label{U5}
\Omega^{(3)}{}_{[ijk}\, \Omega^{(3)}{}_{pqr]}=f_{[ijk} \, f_{pqr]} =0 \quad . 
\e \noindent 
In general, the skewsymmetrisation of two copies of the same Omega tensor
is zero since this corresponds to taking the wedge product of a primitive
$SU(n)$ de Rham cocycle by itself, which is zero because all
these cocyles are represented by odd, ($2m-1$)-forms on the $SU(n)$ 
group manifold. 

\subsection{9-cocycle results}

We used the ninth order Omega tensor to define the fifth order $t$-tensor
$t^{(5)}$, quoting the result \cite{tensors} for it as (\ref{B12}).
This enables us to calculate
\be \label{C35}
\Omega^{(9)}{}_{ijklmpqrs}\Omega^{(9)}{}_{ijklmpqrs} \quad , \e
\noindent to within an overall normalisation constant.
We may use (\ref{B9}) to derive
\bea
\Omega_{ijklmpqrs}\Omega_{ijklmpqrs} & = & 
\Omega_{ijklmpqrs} f^a{}_{[ij} f^b{}_{kl} f^c{}_{pq} f^d{}_{rs]} 
d^{(5)}{}_{(abcdm)}
\nonumber \\
& = & t^{(5)}{}_{abcdm} d^{(5)}{}_{(abcdm)}  \nonumber \\
& = & \lambda(n)   \prod_{r=1}^{4} (n^2-r^2) \quad . \label{C36} \ea \noindent
Here the first line uses the definition (\ref{B6}) and the second one uses
(\ref{B9}). The last line may be evaluated
from the second one using (\ref{B12}) and
the two following results, the first of which is the last equation
of the appendix in \cite{tensors}, and the other is 
much more easily obtained:
\bea 
 d^{(5)}{}_{(abcdm)} d^{(5)}{}_{(abcdm)} & = & \fract{1}{15n^3} (n^2-1)(n^2-4)
(5n^4-96n^2+480) \quad, \label{C37} \\
 d^{(5)}{}_{(abcdm)} \delta_{(ab} d_{cdm)} & = & \fract{1}{5n^2}(n^2-1)(n^2-4)
(3n^2-20) \quad . \label{C38}  
\ea \noindent
Equation
(\ref{C36}) already displays the essential factors anticipated in Sec.
3.5. We confirm its correctness in Sec. 6, using lambda-matrix techniques 
which allow the factor $\lambda(n)$ to be determined, the result having been
given above as (\ref{B12A}).

\section{Selected Derivations}
\subsection{Equations (\ref{C28}), (\ref{C19}) and (\ref{C24}) }

The first result involving $\Omega^{(5)}$ that is not straightforward to derive
from definitions is (\ref{C28}). We develop
\bea
\Omega_{ijklm} \Omega_{ijkpq} & = & \Omega_{mijkl} \Omega_{qijkp} \nonumber \\
 & = & f^a{}_{m[i} f^b{}_{jk]} d_{abl} f^x{}_{q[i} f^y{}_{jk]} d_{xyp} \quad . 
\label{D1} \ea \noindent 
The second set of square brackets can now be 
dropped, leaving behind the sum of three sixth order products of $d$- and 
$f$-tensors to be reduced by trace methods using formulas from the appendix of
\cite{tensors}. Two of the three terms coincide, and nothing more than the 
evaluation of four-fold traces needs to be done. A rough graphical 
representation of any term helps (here and elsewhere) to see the best way to 
use trace formulas. In this vertices correspond in evident manner to
$d$- and $f$-tensors, while closed loops indicate the traces. 
The result comes out 
initially in terms of products of $\delta \delta$ and $dd$ terms. But there is
an identity valid for all $su(n)$ 
(\cite{msw}, eq. (2.10)) which allows the latter to be given in terms
of $\delta \delta$ and $ff$ terms as displayed.

It is not difficult to reduce (\ref{C28}) to confirm 
the correctness of
(\ref{C19})--(\ref{C23}) and hence of (\ref{C24})--(\ref{C27}). 
However, a direct attack on either of the 
latter along the lines just indicated is a good way to get up to speed on 
methods useful in the current study.

\subsection{Equations (\ref{C29}), (\ref{C20}) and (\ref{C25})}

Eq. (\ref{C29}) perhaps discourages such a direct approach as Sec. 4.1
uses, so one adopts a different approach.
This requires, as a
preliminary, the knowledge of (\ref{C25}). Hence we first develop
\bea 
\Omega_{ijklmpq} \Omega_{ijklmpq} & = & \Omega_{ijklmpq}
f^a{}_{[ij} f^b{}_{kl} f^c{}_{mp]} d^{(4)}{}_{(abcq)} \nonumber \\
 & = & t^{(4)}{}_{abcq} d^{(4)}{}_{(abcq)} \nonumber \\
 & = & t^{(4)}{}_{abcq} d_{tab} d_{cqt} \nonumber \\
 & = & \fract{2}{45} n^2 (n^2-9) d_{cqt} d_{cqt} \quad . \label{D2} \ea
\noindent The first line uses the definition (\ref{B6}), from which the square
brackets can be dropped, so that (\ref{B8}) can be used in the second line.
Next the symmetry properties of the $t$-tensors allow the replacement of 
$d^{(4)}$ by one of its terms, whereupon (\ref{B14A}) may be used. 
Since
\be
d_{abc}\, d_{abd} = \fract{(n^2-4)}{n} \, \delta_{cd} \quad, \label{dosd}
\e
eq.  (\ref{C25}) follows.

Returning now to (\ref{C29}), we use that the two terms on the 
right side of (\ref{C28}) are a basis for the vector space of tensors with the
the required symmetry properties. Hence we seek a result of the form
\be \label{D3}
\Omega_{ijklpqr} \Omega_{ijklpst}= b(n)f_{qru} f_{stu}+a(n)
(\delta_{qs}\delta_{rt}-\delta_{qt}\delta_{rs}) \quad . \e \noindent
To determine the coefficients, we must perform contractions with
$\delta_{qs}$ and with $f_{stg}$. As we show below, this gives equations
\bea 
(n^2-2)a(n) +b(n)n & = & \phi_7(n) \nonumber \\
2a(n)+ b(n)n & = & \fract{1}{3} \phi_7(n) \quad , \label{D4} \ea \noindent
which can  be solved to complete the derivation of (\ref{C29}). 
It is easy to get
all the terms here except the one on the right side of the second equation.
A viable starting point is elusive. Consider therefore
\be \label{D5}
\Omega_{ijklpst} f_{gst} f^a{}_{[ij} f^b{}_{kl} f^c{}_{p]q} d^{(4)}{}_{(abcr)}
\quad . \e \noindent
The square brackets can again be dropped.
 This lead us to
\be \label{D6} f_{cpq} d^{(4)}{}_{(abcr)} t^{(4)}{}_{abgp} \e \noindent
using (\ref{B8}). Now, we need to check
\be \label{D7} 
d^{(4)}{}_{(abcr)} \delta_{(ab} \delta_{gp)} =\fract{2}{9} \fract{(n^2-4)}{n}
\delta_{gp} \delta_{cr} \quad , \e \noindent
use (\ref{B14}), and complete the computation 
\be \label{D8}
f_{cpq} d^{(4)}{}_{(abcr)} d_{age} d_{bpe} = \fract{1}{9n^2} (n^2-4)(n^2-8)
f_{qrg}
\quad . \e \noindent
This last one requires use of the four-fold $d$-tensor trace of (A.10)
of \cite{tensors}, some simpler results also found there 
and some patience. Then all the pieces of the calculation have to be 
put together to complete the derivation of (\ref{C29}). One is guided through 
something of a morass to an answer one knows
is right by the fact that both the unknowns $a(n)$ and $b(n)$ 
must contain a factor $(n^2-9)$ that vanishes at $n=3$.

\section{Omega tensors and $su(n)$ lambda-matrices}

\subsection{Antisymmetrised products of $su(n)$ lambda-matrices}

We use the lambda-matrices of ref. \cite{msw}
which are subject to

\be 
\label{E1} {\rm Tr} \, \lambda_i =0\ \quad, \quad 
{\rm Tr} \, \lambda_i \lambda_j = 2\delta_{ij} \quad , \quad 
{\lambda_i}^{\dagger}=\lambda_i  \quad , 
\e
\be 
\label{E2}   
\lambda_i\lambda_j=\fract{2}{n} \delta_{ij} + (d+if)_{ijk} \lambda_k
\quad , \quad d_{ijk}\,\delta_{ij}=0 \quad, 
\e \noindent
and  define totally antisymmetrised products of unit 
weight of lambda-matrices
\be \label{E3}
\lambda_{[ijk \dots s  ]}= \lambda_{[i} \lambda_j \lambda_k \dots \lambda_{s]}
\quad . \e

Simple computations using (\ref{B4}) and (\ref{C2}) lead directly to
\bea
\lambda_{[ijk]} & = & \fract{2}{n}i(f_{ijk}+\Omega_{ijkab} f_{abq} \lambda_q) 
=\fract{2}{n} i f_{ijk}+i f_{s[ij} d_{k]sq} \lambda_q \quad,
\label{E4} \\
\lambda_{[ijkl]} & = & -\Omega_{ijklt} \lambda_t \quad . \label{E5} \ea
\noindent These imply the trace results
\bea 
 {\rm Tr} \,   \lambda_{[ijkl]} & = & 0 \label{E6} \quad, \\
 {\rm Tr} \,   \lambda_{[ijk]l} & = & \fract{4}{n}i \Omega_{ijkab} f_{abl} 
\quad , \label{E7} \ea \noindent
which  may  be contrasted with 
\bea
{\rm Tr} \,   \lambda_{[ijk]} & = & {\rm Tr} \,   \lambda_{i[jk]} 
=2if_{ijk} \quad,
\label{E8} \\
{\rm Tr} \,  \lambda_{[ijkpq]} & = & {\rm Tr} \,  \lambda_{[ijkp]q} 
=-2\Omega_{ijkpq} \quad , \label{E9} 
\ea \noindent
where the first two equalities use the cyclic nature of the trace.
For odd traces as (\ref{E8}), (\ref{E9}) terms related by cyclicity add up,
whereas for even ones they cancel pairwise and indeed completely:
\be \label{E8A} 
{\rm Tr} \,  \lambda_{[i_1 i_2 \dots i_{2s} ]}=0 \quad .\e 
\noindent For the five-fold case one finds
using $\lambda_{[ijkpq]}=\lambda_{[[ijkp]q]}$ and (\ref{C4})
that \be \label{E10}
\lambda_{[ijkpq]}=-\fract{2}{n}\Omega_{ijkpq} - \Omega_{r[ijkp}
d_{q]rt} \lambda_t \quad , \e \noindent
from which (\ref{E9}) can be recovered.
For the six-fold antisymmetrised product of $\lambda$'s we use 
$\lambda_{[ijkpqr]}=\lambda_{[[ijkp][qr]]}$ to deduce
\bea \
\lambda_{[ijkpqr]} & = & -i \Omega_{s[ijkp} f_{qr]t} \lambda_s \lambda_t 
\label{E11} \\
& = & -i \Omega_{ijkpqrs} \lambda_s  \quad , \label{E12} \ea \noindent
and
\be \label{E11A}
{\rm Tr} \, \lambda_{[ijkpq]s}=-2\Omega_{t[ijkp} d_{q]st} \quad . \e \noindent
Here, to derive (\ref{E12}), we used (\ref{E2}) for $\lambda_s \lambda_t$,
$Ad$-invariance (eq. (\ref{C3})) to discard the first term, 
and the following steps to discard another term,
\bea
\Omega_{s[ijkp} f_{qr]t} f_{stm} & = & {f^x}{}_{[ij} {f^y}{}_{kp} {f^t}{}_{qr]}
d_{xys} f_{stu}  \nonumber \\
 & = & {f^x}{}_{[ij} {f^y}{}_{kp} {f^t}{}_{qr]} d_{s(xy} f_{t)ms} =0 
\quad, \label{E13}
\ea \noindent 
upon using (\ref{Y1}). Similar steps, using (\ref{B3}) and (\ref{B5}), 
show how the $d$-term of (\ref{E2}) features in the production 
of (\ref{E12}).

For the seven-fold product we find
\be \label{E12B}  \lambda_{[ijklpqr]}= -\fract{2}{n} i \Omega^{(7)}{}_{ijklpqr}
-i\Omega^{(7)}{}_{s[ijklpq} d_{r]ts} \lambda_t \quad , 
\e 
\noindent 
with the aid of (\ref{C4}) and
\be \label{E12A}  
{\rm Tr} \, \lambda_{[ijkpqrt]}=-2i \Omega^{(7)}{}_{ijkpqrt}
\quad . \e \noindent
Eq. (\ref{E12B}) is 
a natural generalisation of the odd traces (\ref{E4}) and (\ref{E10}), 
and we may infer the result
for the {\it odd} case
\be 
\label{E13A}
\lambda_{[i_1 i_2 \dots i_{2s} k]}= \fract{2}{n} i^s 
\Omega^{(2s+1)}{}_{i_1 i_2 \dots i_{2s} k} +i^s 
\Omega^{(2s+1)}{}_{p [i_1 i_2 \dots i_{2s}}
d_{k]pq} \lambda_q \quad . 
\e \noindent 
Also we may use (\ref{E12A}) and
\be \label{E13B}
{\rm Tr} \, \lambda_{[ijkpqrt]}={\rm Tr} \, \lambda_{[ijkpqr]t}
={\rm Tr} \, \lambda_{[[ijkp][qr]]t} \quad , 
\e \noindent
to check our work by reproducing the recursive identity (\ref{C10}).

Writing the elementary result
\be \label{E14} \lambda_{[ij]}=\fract{1}{2} {[} \lambda_i \, , \,
 \lambda_j {]}=if_{ijk} \lambda_k \quad , \e \noindent
and comparing it also with the even case (\ref{E5}), and (\ref{E12}),
one gets for the antisymmetrised product of an {\it even} 
number of $\lambda$'s the result
\be \label{E15}
\lambda_{[i_1  \dots i_{2s}]}=i^s 
\Omega^{(2s+1)}{}_{i_1 \dots i_{2s} k} \lambda_k  \quad , \e 
\noindent
which implies (\ref{E8A}) and  
\be \label{E16}  
{\rm Tr} \,  \lambda_{[i_1 \dots i_{2s}] k}=2i^s
\Omega^{(2s+1)}{}_{i_1  \dots i_{2s} k} \quad . \e
We note that (\ref{E15}), and in particular (\ref{E5}) and 
(\ref{E12}), provide an explicit realisation of the ($2m-2$)-bracket 
Lie algebras \cite{dApb} for $su(n)$. As mentioned, the coordinates of
the $\Omega^{(2m-1)}$ determine 
the associated  higher order structure constants (above, $s=m-1$), 
and satisfy the GJI (\ref{C8}).

\subsection{General proofs of results for the $\lambda_{[i_1 \cdots i_s]}$}

The above results are general, due to the nature of the 
$\Omega^{(2m-1)}$ tensors as generalised structure constants
(for instance, eq. (\ref{E15}) may be looked at as a consequence 
of Th. 3.1 in \cite{dApb}). However, the above eqs., and in particular 
(\ref{E13A}), were presented on the basis of inspection of a 
modest number of low value cases. It is thus necessary to 
show their general validity, particularly since, as defined
above by eqs. (\ref{B5})--(\ref{B5A}), $\Omega^{(5)}$ 
and $\Omega^{(7)}$ involve $d$-tensors with simple properties
that do not generalise straightforwardly to the $d$-tensors 
involved in the definition of higher $\Omega$-tensors.

Let us look first at (\ref{E15}). We write
\be \label{NN1}
\lambda_{[i_1 j_1 \cdots i_s j_s]}=i^s f^{p_1}{}_{[i_1 j_1} \cdots 
 f^{p_s}{}_{i_s j_s]} \lambda_{(p_1 \cdots p_s)} \quad . 
\e \noindent
If we apply (\ref{E2}) repeatedly, making full use of the 
symmetry properties that are implied by the round brackets, 
we can establish a result of the form
\be 
\label{NN2}
\lambda_{(p_1 \cdots p_s)}={\tilde k}{}_{(p_1 \cdots p_s)t} \lambda_t
+k{}_{(p_1 \cdots p_s\,t)} \quad , 
\e \noindent
where the $k$-tensors are $Ad$-invariant tensors 
with the indicated symmetries. Eq. (\ref{L3.1}) tells us that 
$k$ does not contribute to (\ref{NN1}).
Also ${\tilde k}{}_{(p_1 \cdots p_s)t}$ differs from 
$d^{(s+1)}{}_{(p_1 \cdots p_s)t}$ only by some linear combination of 
non-primitive terms, which, also by (\ref{L3.1}), do not contribute to  
(\ref{NN1}). Further
\be \label{NN3}
{\tilde k}{}_{(p_1 \cdots p_s)t}= \fract{1}{2} {\rm Tr}\,
\lambda_{(p_1 \cdots p_s)t} = \fract{1}{2} {\rm Tr}\, 
\lambda_{(p_1 \cdots p_s t)}
={\tilde k}{}_{(p_1 \cdots p_s t)} \quad , \e \noindent
all of which allows us to replace the lambda-matrix factor of (\ref{NN1}) by 
$d^{(s+1)}{}_{(p_1 \cdots p_s t)} \lambda_t$, so that also
\be \label{NN4}
\lambda_{[i_1 j_1 \cdots i_s j_s]} = \Omega^{(2s+1)}{}_{i_1 j_1 \cdots i_s j_s
t} \lambda_t \quad . \e 
The trace of (\ref{NN4}) now confirms (\ref{E8A}). 

Eq. (\ref{E16}) also follows easily. To obtain 
(\ref{E13A}), we multiply (\ref{NN4}) by $\lambda_k$ and use (\ref{E2}). Then 
(\ref{E13A}) follows directly, after the use of the $Ad$-invariance of 
$\Omega^{(2s+1)}$ to drop the contribution of the $f$-term of (\ref{E2}).

Inspection of (\ref{NN3}) shows that it is tantamount to the statement that, 
in the definition (\ref{B5A}) of $\Omega^{(9)}$, {\it e.g.}, one is, after 
all, allowed to move the right hand round bracket one place to the left. It is
of  interest to see this explicitly, because, amongst other things, a further 
class of identities for $d$-tensors emerges as a by-product. We illustrate 
this for $s=4$. One evaluation of the trace involved leads to 
\be \label{NN5}
\lambda_{(abcd)} =\fract{4}{n^2} \delta_{(ab} \delta_{cd)} 
+\fract{2}{n} d^{(4)}{}_{(abcd)} +\fract{2}{n} d_{(abc}\lambda_{d)}
+\fract{2}{n} \delta_{(ab} d_{cd)y} \lambda_y
+ d^{(5)}{}_{(abcd)y}\lambda_y \quad . 
\e \noindent
The key trace result ({\it cf.} (\ref{NN3}))
\be \label{NN6}
{\rm Tr} \,\lambda_{(abcd)e}= {\rm Tr} \,\lambda_{(abcde)} \quad , \e
\noindent now leads to
\be \label{NN7}
d^{(5)}{}_{(abcde)}=d^{(5)}{}_{(abcd)e}+\fract{1}{n} 
\delta_{(ab} d_{cde)}-\fract{1}{n} \delta_{(ab} d_{cd)e} \quad . \e \noindent
The difference between the two $d^{(5)}$ tensors here, and as in the general
discussion above,  makes no contribution  to
the evaluation of $\lambda_{[i_1 j_1 \cdots i_4 j_4]}$. It follows then that in
the definition (\ref{B5A}), we can replace $d^{(5)}{}_{(abcde)}$ by
$d^{(5)}{}_{(abcd)e}$, which of course has fewer terms.

Another question arises here: how does (\ref{NN6}) relate to (\ref{Y4A})?
To answer, we note that a different way of evaluating the trace gives rise
to
\be \label{NN8}
\lambda_{(abcd)}=\fract{4}{n^2} \delta_{(ab} \delta_{cd)}
+\fract{2}{n} d^{(4)}{}_{(abcd)} 
+\fract{4}{n} \delta_{(ab} d_{cd)y} \lambda_y
+ d^{(5)}{}_{(ab}{}^y{}_{cd)} \lambda_y \quad , \e \noindent
and hence
\be \label{NN9}
d^{(5)}{}_{(abcde)}=d^{(5)}{}_{(ab}{}^e{}_{cd)}-\fract{4}{n} 
\delta_{(ab} d_{cde)}+\fract{4}{n} \delta_{(ab} d_{cd)e} \quad . \e \noindent
Now (\ref{Y4A}) follows obviously form (\ref{NN5}) and (\ref{NN9}).

There is another instructive  way to make the point that the 
three $d^{(5)}$ tensors can be used
interchangeably in the definition (\ref{B5A}) of $\Omega^{(9)}$.
It follows by comparison of  
\bea
\lambda_{[ijklpqrs]} & =& \lambda_{[[ijklpq][rs]]} \nonumber \\
& = & \Omega_{x[ijklpq} f_{rs]y} \lambda_x \lambda_y \nonumber \\
& = & f^a{}_{[ij} f^b{}_{kl} f^c{}_{pq} f^y{}_{rs]} 
{d^{(5)}}_{(abcy)t} \lambda_t \quad , \label{NN10} \ea \noindent
and
\bea
\lambda_{[ijklpqrs]} & =& \lambda_{[[ijk][lpqrs]]} \nonumber \\
& = & f^a{}_{[ij} f^b{}_{kl} f^c{}_{pq} f^y{}_{rs]} 
d^{(5)}{}_{(ab}{}^t{}_{cy)} \lambda_t \quad . \label{NN11} 
\ea 
\noindent
The discussion just given for $\Omega^{(9)}$ generalises naturally for
higher Omega tensors. 

\subsection{Use of the completeness relation for the $su(n)$ lambda-matrices}

We set out from the result well-known for $su(n)$ \cite{msw}
\be \label{P1} \lambda_{i \, ab} \lambda_{i \, cd}
=2\delta_{ad} \delta_{cb}-\fract{2}{n} \delta_{ab} \delta_{cd} \quad , \e
\noindent and note also its consequences
\bea  -if_{ijk} \lambda_{j \, ab} \lambda_{k \, cd} & = & \lambda_{i \, ad}
 \delta_{cb}-\lambda_{i \, cb} \delta_{ad} \label{P2} \\
     d_{ijk} \lambda_{j \, ab} \lambda_{k \, cd} & = & \lambda_{i \, ad} 
\delta_{bc}+\lambda_{i \, cb} \delta_{ad} 
-\fract{2}{n} \left(  \lambda_{i \, ab} \delta_{cd} +
  \lambda_{i \, cd} \delta_{ab} \right) \quad . \label{P3} \ea 
\noindent From (\ref{P1}) we may compute
\bea \lambda_{[ij] \, ab} \lambda_{[ij] \, cd}
 & = & -n \lambda_{i \, ab} \lambda_{i \, cd} \label{P4} \\
\lambda_{[ijk] \, ab} \lambda_{[ijk] \, cd}
 & = & -\fract{2}{3} (n^2-4) \lambda_{i \, ab} \lambda_{i \, cd} 
       -\fract{4}{n} (n^2-1) \delta_{ab} \delta_{cd} \label{P5} \\
\lambda_{[ijkl] \, ab} \lambda_{[ijkl] \, cd}
 & = & \fract{n}{3} (n^2-4)  \lambda_{i \, ab} \lambda_{i \, cd} \label{P6} \\
\lambda_{[ijklm] \, ab} \lambda_{[ijklm] \, cd}
 & = & \fract{2}{15} (n^2-4)(n^2-6) \lambda_{i \, ab} \lambda_{i \, cd} 
+\fract{4}{3n} (n^2-1)(n^2-4) \delta_{ab} \delta_{cd} \
\label{P6A} \; , \ea 
\noindent and so on. One can make checks on these results by 
putting $b=c$ to reach
\bea 
\lambda_i \lambda_i & = & \fract{2}{n} (n^2-1) I \label{P7} \\
\lambda_{[ij]} \lambda_{[ij]} & = & -2(n^2-1) I \label{P8} \\
\lambda_{[ijk]} \lambda_{[ijk]} & = & -\fract{4}{3n} (n^2-1)^2  I \label{P9} \\
\lambda_{[ijkl]} \lambda_{[ijkl]} & = & \fract{2}{3} (n^2-1)(n^2-4) I 
\label{P10} \\
\lambda_{[ijklm]} \lambda_{[ijklm]} & = & \fract{4}{15n} (n^2-1)^2 (n^2-4) I 
\label{P10A} \quad. \ea
\noindent 
These results are of use themselves and may be verified by other 
means. It is tempting to speculate on the nature of results beyond (\ref{P6A}),
but it gets increasingly hard to compute directly the $n$-dependences.
Use of ${\rm Tr} \, I=n$ yields obvious trace formulas. 

For the purpose, central to the aims of this paper, of computing the
quantities $(\Omega^{(2m+1)})^2$ explicitly in closed form, it is enough to 
analyse the traced analogues of (\ref{P4})--(\ref{P6A}), obtained by
putting $b=c$ and $d=a$. For this analysis, one of the approaches available 
employs another
set of lemmas that follow from (\ref{P1}). For any
$n$-dimensional matrix $M$, eq. (\ref{P1}) gives
\be \label{P11} (\lambda_i M \lambda_i)_{ab} = 2\delta_{ab} {\rm Tr} \, M
-\fract{2}{n} M_{ab} \quad . \e \noindent
This provides us with a method for obtaining the results
\bea 
\lambda_i \lambda_j \lambda_i & = & -\fract{2}{n} \lambda_j \label{P12} \\
\lambda_i \lambda_{[jk]} \lambda_i & = & -\fract{2}{n} \lambda_{[jk]} 
\label{P13} \\
\lambda_i \lambda_{[jkl]} \lambda_i & = & 4if_{jkl}
-\fract{2}{n} \lambda_{[jkl]} \label{P14} \\
\lambda_i \lambda_{[jklpq]} \lambda_i & = & -4\Omega_{jklpq}
-\fract{2}{n} \lambda_{[jklpq]} \label{P15} \\
\lambda_i \lambda_{[i_1 j_1 \dots i_m j_m k]} \lambda_i & = & 4i^m
{\Omega^{(2m+1)}}_{i_1 j_1 \dots i_m j_m k }
-\fract{2}{n} \lambda_{[i_1 j_1 \dots i_m j_m k]} \label{P16} \\
\lambda_i \lambda_{[i_1 j_1 \dots i_m j_m]} \lambda_i & = & -\fract{2}{n} 
\lambda_{[i_1 j_1 \dots i_m j_m]} \quad . \label{P17} \ea 
\noindent The last result follows from (\ref{P11}) because of (\ref{E8A}).

We note here further simple results that may help streamline larger tasks,
for example one approach to the proof of (\ref{P7})--(\ref{P10A}):
\bea
\lambda_i \lambda_{[ij]} & = & n \lambda_j \label{P18} \\
\lambda_i \lambda_{[ijk]} & = & \fract{2}{3} \fract{(n^2-1)}{n} \lambda_{[jk]}
\label{P19} \\
\lambda_{[ij]} \lambda_{[ijk]} & = & - \fract{2}{3} (n^2-1) \lambda_k 
\label{P20}  \\
\lambda_i \lambda_{[ijkl]} & = & \fract{n}{2} \lambda_{[jkl]}-i f_{jkl} 
\label{P20A} \\
\lambda_{[ij]} \lambda_{[ijkl]} & = & - \fract{1}{3} (n^2-4) \lambda_{[kl]}
\label{P21} \\
\lambda_{[ijk]} \lambda_{[ijkl]} & = & - \fract{n}{3} (n^2-4) \lambda_l
\label{P22} \\
\lambda_i \lambda_{[ijklm]} & = & \fract{2}{5n} (n^2-1) \lambda_{[jklm]}
-\fract{4}{5} i f_{{[}jkl} \lambda_{m{]}} \label{P23} \\
\lambda_{[ij]} \lambda_{[ijklm]} & = & -\fract{1}{5} (n^2-4) \lambda_{[klm]}
 \label{P24} \\
\lambda_{[ijk]} \lambda_{[ijklm]} & = & -\fract{2}{15n} (n^2-1)(n^2-4) 
\lambda_{[lm]} \label{P25} \\
\lambda_{[ijkl]} \lambda_{[ijklm]} & = & \fract{2}{15} (n^2-1)(n^2-4) 
\lambda_m \quad . \label{P26} 
\ea 
Hermitian conjugation gives results such as
\be \label{P27} \lambda_{[ji]} \lambda_i= n \lambda_j \quad . \e
\noindent

Inspection of (\ref{P18}), (\ref{P19}), (\ref{P20A}) and (\ref{P23}) suggests
the general result 
\be \label{P101}
{\rm Tr}\,\lambda_i \lambda_{[i i_2 \dots i_s]}=0 \quad 
(s \quad {\rm even \quad  or \quad  odd})
\quad, \e \noindent
which is easily proved using the results of Sec. 5.2.
If $s$ is {\it even} we find, using (\ref{E15}), 
${\rm Tr}(\lambda_i\lambda_{[ii_2\dots i_s]})\sim 
{\rm Tr}(\lambda_i\Omega_{ii_2\dots i_s k}\lambda_k)=
2\Omega_{ii_2\dots i_si}=0$. If $s$ is {\it odd},
${\rm Tr}(\lambda_i\lambda_{[ii_2\dots i_s]})= 
{\rm Tr}(\lambda_i\lambda_{[i[i_2\dots i_s]]}) \\
\sim{\rm Tr}(\lambda_i\lambda_{[i}\lambda_{k]})\, \Omega_{i_2\dots i_s k}=0$.

 Some of the principal results 
to be derived below require as input more trace results.
First, and in agreement with results displayed above, we expect 
\be \label{P104}
{\rm Tr}\,(\lambda_{[ij]} \lambda_{[i j i_3 \dots i_{2s}]})=0 \quad . \e 
\noindent 
A typical proof here, using the methods of Sec. 5.2, is
\be 
\label{P105}
{\rm Tr}(\lambda_{[ij]} \lambda_{[ijklpq]})=
{\rm Tr}(\lambda_{[ij]} (-i\Omega_{ijklpqr} \lambda_r))
=2\Omega_{ijklpqr}f_{ijr} =0 \quad , 
\e \noindent 
upon use of (\ref{K5}). The analogues to (\ref{P104}) for $s=4$ and 
$s=5$ of (\ref{P104}) however depend on (\ref{K6}) and (\ref{K7}), results 
which remain unproved until the methods of Sec. 8 can be called upon.
Second
\be \label{P106}
{\rm Tr}\,(\lambda_{[ijk]} \lambda_{[ijk  i_4 \dots i_{2s}]})=0 \quad . \e 
\noindent 
A typical proof, here for $s=3$, is
\bea
{\rm Tr}\,(\lambda_{[ijk]} \lambda_{[ijklpq]}) & = & {\rm Tr}\,(\lambda_{[ijk]} 
(-i\Omega_{ijklpqr} \lambda_r)) \nonumber \\
& = & \Omega_{ijklpqr} \fract{4}{n} \Omega_{ijkab} f_{abr} \nonumber \\
& = & 2\Omega_{ijklpqr} f_{u[ij} d_{k]ur} =0 \quad ,
\label{P107} 
\ea \noindent
where (\ref{E12}), (\ref{E7}) and (\ref{C14}) have been used.
The last equality follows from the total symmetry  of $d^{(3)}$
and the antisymmentry of $\Omega^{(7)}$.

\subsection{Further trace results}

We are interested here in trace results of the type
\be \label{P28}
 {\rm Tr} \, (\lambda_{[ij \dots s]} \lambda_{[ij \dots s]}) \quad . \e
\noindent 
Even traces of this sort are of primary interest in virtue of their 
relationship to $(\Omega^{(2s+1)})^2$. For example, for $s=2$,
\bea 
4 (\Omega^{(5)})^2 & = & {\rm Tr} \, \lambda_{[ijkpq]} \;{\rm Tr} \, 
\lambda_{[ijkpq]} \nonumber \\
 & = &  {\rm Tr} \, \lambda_i \lambda_{[jkpq]} \;{\rm Tr} \, \lambda_i
\lambda_{[jkpq]} \nonumber\\
 & = & 2 {\rm Tr} \, \left( \lambda_{[jkpq]} \lambda_{[jkpq]} \right) \quad , 
\label{P29} \ea
where (\ref{P1}) and (\ref{E6}) have been used. However, 
proceeding recursively for higher
$s$ brings the odd traces into the picture. We have two
approaches to either even or odd traces, and one works better for the odd and 
the other for the even traces. We begin with the even traces for which we have
a nice general result.
We note a generalisation of results embedded in the previous subsections:
\be \label {P30}
{\rm Tr} \, (\lambda_{[i_1 \dots i_{2s} k]} 
            \lambda_{[i_1 \dots i_{2s} k]}) 
=\fract{2}{(2s+1)} \; \fract{(n^2-1)}{n}
{\rm Tr} \, (\lambda_{[i_1 \dots i_{2s} ]} 
            \lambda_{[i_1 \dots i_{2s} ]}) 
\quad . \e
\noindent 
A brief look at the case $s=3$ will indicate clearly that this 
result is valid in general. Thus we write
\bea
{\rm Tr} \, (\lambda_{[ijklpqr]} \lambda_i \lambda_j \lambda_k \lambda_l
 \lambda_p  \lambda_q \lambda_r) & = & 
\fract{1}{7}  {\rm Tr}  \, (\lambda_i \lambda_{[jklpqr]}      
\lambda_i \lambda_{[jklpqr]}) \nonumber \\
& + &   \fract{1}{7} {\rm Tr} \, (\lambda_j \lambda_{[klpqri]}      
\lambda_i \lambda_j \lambda_{[klpqr]}) + \dots  \quad . \label{P31} \ea
\noindent
Now we use the cyclic property of the trace to justify 
the use of results of the 
type (\ref{P16}) and (\ref{P17}) in the first six terms, and of
(\ref{P7}) to the seventh. One can see from (\ref{E8A}) the Omega tensor terms
of (\ref{P16}) do not contribute (which is why this approach is better for the 
odd traces than for the even ones), and then it is easy to see 
that, after taking due care of the signs of the first six terms, everything 
cancels except the contribution of the
seventh term of (\ref{P31}), which gives the right side of (\ref{P30}) at
$s=3$. 

Reduction of the right side of (\ref{P30}) is much harder because
the same approach brings in the Omega tensor pieces of (\ref{P14}), 
(\ref{P16}) {\it etc.}, non-trivially. This caused us to adopt a related 
but distinct approach to such traces in the next section, although the approach
just followed does work, but rather less well. To say enough to allow a 
comparison of methods to be made, let us refer back to (\ref{P29}). We may 
drop the second set of square brackets and reinsert others judiciously in 
suitable places whenever this is allowed by existing antisymmetries.
Then the development of the first set of square brackets yields
\bea
8(\Omega^{(5)})^2 & = & {\rm Tr} ( \lambda_j \lambda_{[kpq]}
 \lambda_j \lambda_{[kpq]}) \nonumber \\
& - &   {\rm Tr} ( \lambda_k \lambda_{[pqj]} \lambda_j \lambda_k 
\lambda_{[pq]}) \nonumber \\
& + &  {\rm Tr} ( \lambda_p \lambda_{[qjk]} \lambda_j \lambda_k \lambda_p 
\lambda_q) \nonumber \\
& - &   {\rm Tr} ( \lambda_q \lambda_{[jkp]} \lambda_j \lambda_k 
\lambda_p \lambda_q)   \quad . \label{P32} \ea
\noindent
The cyclic property of the trace now allows use of (\ref{P14}),(\ref{P13}), 
(\ref{P12}) and (\ref{P7}), in that order so that after cancellations, 
we obtain 
\be \label{P33}
8(\Omega^{(5)})^2 ={\rm Tr} \, \lambda_{[kpq]}4if_{kpq}
+\left( \fract{2}{n} +\fract{2}{n}(n^2-1) \right) {\rm Tr} ( 
\lambda_{[kpq]} \lambda_k \lambda_p \lambda_q)  \quad . \e \noindent
Now use of (\ref{E8}) and (\ref{P9}) leads directly to the answer obtained 
before: (\ref{C24}) with (\ref{C26}). The higher order even traces get 
successively harder in this approach, but we will see a comparable 
increase in the
price associated with passing to higher $s$ is present also 
in our favoured method of Sec. 6.

\section{The recursion relations for the $(\Omega^{(2m-1)})^2$}

We illustrate the general approach by reference to the case
$m=5$. Since, by eq. (\ref{E16}),
\be 
\label{S1} 2 \Omega_{ijklpqrst}={\rm Tr} \, \lambda_{[ijklpqrs]t} 
\quad , \e
\noindent 
we may write
\bea 
4 (\Omega^{(9)})^2 = 4 \Omega_{ijklpqrst} \, \Omega_{ijklpqrst}
 & = & {\rm Tr} \, \left( \lambda_{[ijklpqrs]} \lambda_t \right) \,
{\rm Tr} \, \left( \lambda_{[ijklpqrs]} \lambda_t \right) \nonumber \\
 & = & 2 {\rm Tr} \, \left( \lambda_{[ijklpqrs]} \lambda_{[ijklpqrs]} \right)
\; . \label{S2} \ea \noindent Here we have used 
(\ref{P1}) and the trace
result (\ref{E8A}).  The key steps now follow. We can remove the first set
of square brackets completely and then reinsert them round the indices 
$jklpqrs$. Then we expand the second set of square brackets to expose, in each 
of the eight terms that thereby arise, the matrix $\lambda_i$:
\bea
16 (\Omega^{(9)})^2= \lambda_{i \, ab} \lambda_{[jklpqrs] \,bc} \Big(
 \; \delta_{cd} \lambda_{i \, de} \lambda_{[jklpqrs] \, ea} 
 & - & \lambda_{s \, cd} \lambda_{i \, de} \lambda_{[jklpqr] \, ea} 
\nonumber \\
+ \lambda_{[rs] \, cd} \lambda_{i \, de} \lambda_{[jklpq] \, ea} 
 & - & \lambda_{[qrs] \, cd} \lambda_{i \, de} \lambda_{[jklp] \, ea} 
\nonumber \\
+ \lambda_{[pqrs] \, cd} \lambda_{i \, de} \lambda_{[jkl] \, ea} 
 & - & \lambda_{[lpqrs] \, cd} \lambda_{i \, de} \lambda_{[jk] \, ea} 
\nonumber \\
+ \lambda_{[klpqrs] \, cd} \lambda_{i \, de} \lambda_{j \, ea} 
 & - & \lambda_{[jklpqrs] \, cd} \lambda_{i \, de} \delta_{ea} \; \Big) \quad ,
\label{S3} 
\ea 
\noindent 
where $a,b,\dots,e=1,\dots,n$ are matrix element indices, 
$\lambda_{i\,ab}\equiv {(\lambda_i)}{}_{ab}$.
Now we may use (\ref{P1}) once more. The second term of (\ref{P1}) 
gives zero contribution, or rather its contributions to the eight 
terms of (\ref{S3}) cancel pairwise. Turning next to the contributions 
that come from the first term of (\ref{P1}), we see
the second, fourth, sixth and seventh terms of (\ref{S3}) vanish because of 
trace results such as (\ref{E8A}). The first term gives
\be \label{S4} 
 2 {\rm Tr} \, \left( \lambda_{[jklpqrs]} \right) \,
{\rm Tr} \, \left( \lambda_{[jklpqrs]} \right)= -8 (\Omega^{(7)})^2 \quad ,
\e \noindent by steps like those that yielded (\ref{S2}). The eighth 
term gives 
\bea 
-2 {\rm Tr} \, \left( \lambda_{[jklpqrs]} \lambda_{[jklpqrs]} \right) 
{\rm Tr} \, I_n
 & = & (-2) \fract{2}{7} \fract{(n^2-1)}{n} {\rm Tr} \, \left( 
\lambda_{[jklpqr]}  \lambda_{[jklpqr]} \right) n \nonumber \\
 & = & - \fract{8}{7} (n^2-1) (\Omega^{(7)})^2 \quad , \label{S5} \ea 
\noindent where the result (\ref{P30}) has been used. There thus remains 
to  be treated a set
of two terms, one each from the third and fifth lines of (\ref{S3}). We 
next display the two terms in question together with the results of 
evaluating them 
\bea 
2 {\rm Tr} ( \lambda_{[jklpqrs]} \lambda_{[rs]} ) {\rm Tr}  
(\lambda_{[jklpq]}) & = & -8.\fract{5}{7} (\Omega^{(7)})^2  \label{S6}\quad, \\
2 {\rm Tr} ( \lambda_{[jklpqrs]} \lambda_{[pqrs]} ) {\rm Tr} ( 
\lambda_{[jkl]}) & = & -8.\fract{3}{7} (\Omega^{(7)})^2  \quad . \label{S7} 
\ea \noindent Proofs of (\ref{S6}) and (\ref{S7}) are given in below
in Sec. 7. We may now collect the contributions (\ref{S4})--(\ref{S7}) to 
produce the final answer
\be \label{S8} 
(\Omega^{(9)})^2 = \fract{1}{14} (n^2-16)(\Omega^{(7)})^2 \quad . \e
\noindent We have presented this calculation in detail because every aspect 
of it works for higher cases in almost exactly the same fashion. 
The main difference for the case $m=5$ of $(\Omega^{(11)})^2$ is that there are
now three terms in the  set of
terms that arise in the same way as did (\ref{S6}) and (\ref{S7}), namely   
\bea 
2 {\rm Tr} ( \lambda_{[jklpqrsuv]} \lambda_{[uv]})  {\rm Tr}\, 
\lambda_{[jklpqrs]} & = & -8.\fract{7}{9} (\Omega^{(9)})^2  \label{S9} \\
2 {\rm Tr} ( \lambda_{[jklpqrsuv]} \lambda_{[rsuv]}) \, {\rm Tr} \, 
\lambda_{[jklpq]} & = & -8.\fract{5}{9} (\Omega^{(9)})^2  \label{S10} \\
2 {\rm Tr} ( \lambda_{[jklpqrsuv]} \lambda_{[pqrsuv]}) \, {\rm Tr} \, 
\lambda_{[jkl]} & = & -8.\fract{3}{9} (\Omega^{(9)})^2  \quad . \label{S11} \ea
\noindent With the aid of these results, proved below in Sec 7.1, we 
reach the $m=5$ analogue of (\ref{S8})
\be \label{S12} 
(\Omega^{(11)})^2 = \fract{2}{45} (n^2-25)(\Omega^{(7)})^2 
\quad . \e

Indeed it is possible to infer the general result relating the squares 
of the $(2s-1)$- and $(2s+1)$-cocycles associated with the 
Racah-Casimir operators of order $s$ and $(s+1)$:
\be \label{S13} 
(\Omega^{(2s+1)})^2 = \fract{4}{2s(2s-1)} (n^2-s^2)(\Omega^{(2s-1)})^2 
\quad , \e
\noindent and hence
\be \label{S14}
(\Omega^{(2m-1)})^2 =\frac{2^{2m-3}\, n}{(2m-2)!}\, \prod_{r=1}^{m-1} (n^2-r^2)
\quad . \e
The last two results  show in full the expected factors that 
force the absence of the 
$su(n)$-algebra cocycle/Omega tensor 
$\Omega^{(2m-1)}$ whenever $m>n$. Indeed, the last factor in (\ref{S14}) is 
$(n^2-(m-1)^2)$ and hence $(\Omega^{(2m-1)})^2=0$ whenever $n<m$.
These results are also
crucial in the discussion \cite{dAMcas} of 
Racah-Casimir operators, their eigenvalues and of generalised Dynkin indices 
for $su(n)$.

\section{Proof of results like (\ref{S6})-(\ref{S11})}
 
We begin with the simplest result (\ref{S6}) for which we develop
\bea
2 {\rm Tr}( \lambda_{[jklpqrs]} \lambda_{[rs]}) {\rm Tr} \, \lambda_{[jklpq]}
& = & -4i {\rm Tr} \, [\left( -\fract{2}{n} i \Omega_{jklpqrs}
-i \Omega_{t[jklpqr} d_{s]tx} \lambda_x \right) i f_{rsy} 
\lambda_y] \Omega_{jklpq}
\nonumber \\
& = & -8 \Omega_{t[jklpqr} d_{s]tx} f_{rsx} \Omega_{jklpq}
\nonumber \\
& = & -8.\fract{5}{7} \Omega_{tklpqrs} d_{jtx} f_{rsx} \Omega_{jklpq}
\nonumber \\
& = & -\fract{40}{7} \Omega_{tklpqrs} f_{kla} f_{pqb} d_{abj} f_{rsx} d_{jtx}
\nonumber \\
& = & -\fract{40}{7} \Omega_{tklpqrs} f_{kla} f_{pqb} f_{rsx} 
{d^{(4)}}_{(tabx)} \nonumber \\
& = & -\fract{40}{7} \Omega_{tklpqrs} \Omega_{tklpqrs} 
\quad , \label{T1} 
\ea \noindent
where we used (\ref{E12B}) and (\ref{E10}) in the first line.
Next, by opening up the square brackets, we find seven terms of which two  
vanish upon  use of (\ref{K4}), while the remaining five are seen to be 
equal after relabelling. This accounts for the fraction that appears in the 
third line. In the fourth line we have used the definition of $\Omega_{jklpq}$,
which sets the scene for using the definition (\ref{B5})
of $ \Omega_{tklpqrs}$ to reach the last line. It may be noted that it is the 
first Omega tensor which allows the required symmetries to be implied for
the remaining factors in order to build the second Omega tensor. 

It should suffice to illustrate things fully to sketch the proof of the most
complicated member of the set of results (\ref{S6})--(\ref{S11}). 
This requires the $su(n)$ relation:
\be \label{T2} 
{\rm Tr}  (\lambda_t \lambda_{(abc)})=
{\rm Tr}  \lambda_{(tabc)}=
\fract{4}{n} \delta_{t(a} \delta_{bc)}
+2d_{st(a} d_{bc)s} \quad , \e 
\noindent 
which also follows from (\ref{NN5}). Then, putting in some 
brackets, we get
\bea
2 {\rm Tr} ( \lambda_{[jklpqrsuv]} \lambda_{[[pq][rs][uv]]}) \, {\rm Tr} \, 
\lambda_{[jkl]} & = & 4 {\rm Tr} \, \Omega_{w[jklpqrsu} d_{v]wt} \lambda_t
f_{pqa} f_{rsb} f_{uvc} \lambda_a \lambda_b \lambda_c f_{jkl} \nonumber \\
 & = & 8 \Omega_{w[jklpqrsu} d_{v]wt} 
f_{pqa} f_{rsb} f_{uvc} f_{jkl} d_{ht(a}\, d_{bc)h} \nonumber \\
& = & 8.\fract{3}{9}  \Omega_{wklpqrsuv} d_{jwt} f_{klj}
f_{pqa} f_{rsb} f_{uvc} d_{ht(a} d_{bc)h} \nonumber \\
& = & \fract{24}{9} \Omega_{wklpqrsuv} f_{klj} f_{pqa} f_{rsb} f_{uvc} 
{d^{(5)}}_{(abcj)w} \nonumber \\
& = & \fract{24}{9} \Omega_{wklpqrsuv} \Omega_{wklpqrsuv} \quad . 
\label{T3} \ea
\noindent Here the steps can be seen to be similar to steps already used.
In virtue of the work of Sec. 3.2, the first Omega tensor
provides enough  antisymmetry properties to justify the use of (\ref{B6})
to identify the second one. The fraction in the third line follows opening 
up of square brackets to expose nine terms, of which six vanish because of 
(\ref{K6}), {\it i.e.} 
\be \label{T4} \Omega^{(9)}{}_{wjklpqrsu} f_{jkl} =0 \quad, \e \noindent 
and the remaining three are equal. The first term of  
(\ref{T2}) fails to contribute to line two of (\ref{T3})
because of Jacobi identities that also rely 
on the antisymmetry properties provided by the first Omega tensor. 

It can be seen that as one goes, notionally, to higher $m$ all the same 
patterns persist. Although this may require results beyond those explicitly
provided here, no problems should be encountered 
in finding these, the generalisations of (\ref{K6}) being given in Sec.8.
We remark also that the 
coefficients that appear on the right side of (\ref{S6}) and (\ref{S7}),
and on the right side of (\ref{S9})--(\ref{S11}) also conform to a rather 
obvious pattern, which affords a check on the work, and is instrumental in
producing the crucial $(n^2-s^2)$ factors of the 
recursion relations (\ref{S13}).

\section{Proof of (\ref{K7})}

We first prove here that
\be \label{X101}
\Omega^{(11)}{}_{ijklpqrstuv} f_{tuv} =0 \quad .
\e \noindent This is a critical case because it is the simplest one of the type
discussed in Sec.3.4 in which 
there is a non-trivial invariant totally antisymmetric tensor of the same
order as the right side of the identity to be proved, namely the tensor
$\tilde{\Omega}{}^{(8)}$ of (\ref{U2}).
In this case, moreover, the methods of Sec. 3.4 do not offer a viable approach.
The method of proof to be given for (\ref{X101})
extends straightforwardly to its analogue for
$\Omega^{(13)}$. But then we meet a further critical case 
\be \label{X102}
\Omega^{(15)}{}_{ijklpqrstuvwxyz} f_{xyz} =0 \quad .
\e \noindent This case is critical because it is the simplest one in which
there is a non-trivial invariant totally antisymmetric tensor of the same order
as $\Omega^{(15)}$ itself, the tensor $\tilde{\Omega}{}^{(15)}$ of (\ref{U3}).
However there is no obstacle to extending to this case the method of proof 
to be given for (\ref{X101}).
 
To prove (\ref{X101}), we start with
\bea 
4 \Omega^{(11)}{}_{ijklpqrstuv} f_{tuv} & = & 
{\rm Tr} \, \lambda_{[ijklpqrstuv]}
{\rm Tr} \, \lambda_{[tuv]} \nonumber \\
& = & {\rm Tr} \, \lambda_{[ijklpqrstu]v} {\rm Tr} \, \lambda_{[tu]v} 
\nonumber \\
& = & 2 {\rm Tr} \, \lambda_{[ijklpqrstu]}\lambda_{[tu]} \quad , 
\label{X1} \ea
\noindent using now familiar steps.

To make progress with computing the right side of (\ref{X1}),
we drop the second set of square brackets and open 
out the other set to expose $\lambda_u$ in each of its ten terms. This serves
to enable a second use of (\ref{P1}):
\bea 
( \quad \lambda_{[ijklpqrst] \, ab} \delta_{dc}
& - & \lambda_{[jklpqrst \, ab} \lambda_{i]\, dc} \nonumber \\
+ \lambda_{[klpqrst \, ab} \lambda_{ij]\, dc}
& - & \lambda_{[lpqrst \, ab} \lambda_{ijk]\, dc} \nonumber \\
+ \lambda_{[pqrst \, ab} \lambda_{ijkl]\, dc}
& - & \lambda_{[qrst \, ab} \lambda_{ijklp]\, dc} \nonumber \\
+ \lambda_{[rst \, ab} \lambda_{ijklpq]\, dc}
& - & \lambda_{[st \, ab} \lambda_{ijklpqr]\, dc} \nonumber \\
+ \lambda_{[t \, ab} \lambda_{ijklpqrs]\, dc}
& - & \delta_{ab} \lambda_{[ijklpqrst]\, dc} \; \; ) \quad \fract{1}{5}
\lambda_{u \, bd} 
\lambda_{t \, ce} \lambda_{u \, ea}  \; , \label{X2} \ea \noindent
where the labels $a,\dots,e=1,\dots,n$ are matrix element indices, 
hence unaffected by antisymmetrisation.

It is easy to check that all ten contributions from
the second term of (\ref{P1}) cancel pairwise. The ten contributions from the 
first term of (\ref{P1}) then are 
$\fract{2}{5}$ times
\bea
{\rm Tr} \, \lambda_{[ijklpqrst]} {\rm Tr} \, \lambda_t
& - &  {\rm Tr} \, \lambda_{[jklpqrst} {\rm Tr} \, \lambda_{i]t}
\nonumber \\
+   {\rm Tr} \, \lambda_{[klpqrst} {\rm Tr} \, \lambda_{ij]t}
& - &  {\rm Tr} \, \lambda_{[lpqrst]} {\rm Tr} \, \lambda_{ijk]t}
\nonumber \\
+   {\rm Tr} \, \lambda_{[pqrst} {\rm Tr} \, \lambda_{ijkl]t}
& - &  {\rm Tr} \, \lambda_{[qrst]} {\rm Tr} \, \lambda_{ijklp]t}
\nonumber \\
+   {\rm Tr} \, \lambda_{[rst} {\rm Tr} \, \lambda_{ijklpq]t}
& - &  {\rm Tr} \, \lambda_{[st} {\rm Tr} \, \lambda_{ijklpqr]t}
\nonumber \\
+  {\rm Tr} \, \lambda_{[t} {\rm Tr} \, \lambda_{ijklpqrs]t}
& - &  n {\rm Tr} \, (\lambda_{[ijklpqrst]} \lambda_t)
\quad . \label{X3} \ea \noindent 
Terms 1 and 9 here are zero trivially, terms 2, 4, 6, 8  are zero 
using (\ref{E8A}). Also term 10 is zero
by (\ref{P101}). This leaves
terms 3, 5 and 7. Terms 3, 5 and 7 are, to within a common factor, given by
\bea
& {} & \Omega^{(7)}{}_{[klpqrst} \, f_{ij]t} \label{X4} \\
& {} & \Omega^{(5)}{}_{[pqrst} \, \Omega^{(5)}{}_{ijkl]t} \label{X5} \\
& {} & f_{[rst} \Omega^{(7)}{}_{ijklpq]t} \quad . \label{X6} 
\ea \noindent
The terms (\ref{X4}), (\ref{X6}) are zero since they
are the result of extending the antisymmetrisation of expressions that 
are already zero by $Ad$-invariance, {\it cf.} (\ref{C4}).
Similarly the term (\ref{X5}) is zero by (\ref{C8}). 

\vskip 1.5cm
{\bf Acknowledgements}. This work was partly supported by the
DGICYT, Spain ($\#$PB 96-0756) and PPARC, UK. One of the authors
(JA) wishes to thank the theory group at Imperial College,
London, for their hospitality during the last stages of this paper.

\end{document}